\documentstyle[prd,aps]{revtex}
\begin{document}
\draft
\def\lsim{\mathrel{\lower2.5pt\vbox{\lineskip=0pt\baselineskip=0pt
           \hbox{$<$}\hbox{$\sim$}}}}
\def\gsim{\mathrel{\lower2.5pt\vbox{\lineskip=0pt\baselineskip=0pt
           \hbox{$>$}\hbox{$\sim$}}}}

%
%
\twocolumn[\hsize\textwidth\columnwidth\hsize\csname
@twocolumnfalse\endcsname

\preprint{SUSSEX-AST-95/11-5, IEM-FT-95/120, astro-ph/9511078}
\title{Open Inflation\thanks{To appear in the Proceedings of
TAUP '95, Toledo, September 1995.}}
\author{Juan Garc\'\i a-Bellido}
\address{Astronomy Centre, University of Sussex, Falmer, Brighton BN1
9QH,\ United Kingdom}
\date{November 16, 1995}
\maketitle
\begin{abstract}
Open Inflation has recently been suggested as a possible way out of
the age crisis caused by observations of a large rate of expansion of
the universe, in conflict with the existence of very old globular
clusters. It proposes that our local patch of the universe originated
in a quantum tunneling event, with the formation of a single bubble
within which our universe inflated to almost flatness. I review the
different models proposed together with their predictions for the
amplitude of temperature anisotropies in the cosmic microwave
background.
\end{abstract}
\pacs{98.80.Cq \hspace*{3cm} Sussex preprint SUSSEX-AST 95/11-5,
astro-ph/9511078}

\vskip2pc]

\vspace{1cm}
\section{INTRODUCTION}

Until recently, one of the most robust predictions of inflation was
the extreme flatness of our local patch of the universe. However, in
the last few months there has been a lot of excitement about the
possibility of producing an open universe from
inflation~\cite{BGT,STY,LM}. An open universe could resolve the age
crisis caused by the observations of a relatively large Hubble
constant, $H_0 = 69 \pm 8$ km/s/Mpc, which corresponds (for $\Omega_0
= 1$ and $\Lambda = 0$) to a very small age of the universe, $t_0 =
9.5 \pm 1.1$ Gyr~\cite{Nature}, in conflict with the age of globular
clusters, $12 - 17$ Gyr~\cite{GlobClus}. An alternative solution could
be the introduction of a non-zero cosmological constant $\Lambda$
which could accommodate both a flat and old universe with a large
expansion rate, but there still remains the question of why $\Lambda$
is so small. With `open inflation' it is possible to produce a very
homogeneous open universe, without the need to introduce a
cosmological constant.\footnote{To appear in the Proceedings of
TAUP '95, Toledo, September 1995.}

In standard inflation the homogeneity and flatness problems of the hot
big bang cosmology are intimately related and it is not possible to
relax one (flatness) without affecting the other
(homogeneity)~\cite{KTF}. During inflation both curvature and
inhomogeneities are stretched away. The present value of
$\Omega=\rho/\rho_c$ depends on the number of $e$-folds $N_e$ during
inflation, $|1-\Omega_0| \simeq 10^{57}\,e^{-2 N_e}$, and thus a large
expansion, $N_e \gg 65$, produces an exponentially flat universe, with
the size of the homogeneous patch $L_0$ much greater than the present
horizon, $H_0^{-1}$. In order to produce an open universe, $\Omega_0
\lsim 1$, one requires $N_e \lsim 65$, which means that $L_0 \simeq
H_0^{-1}$. On the other hand, assuming that the density perturbations
at large scales $L\geq L_0$, with density contrast $\delta_L \simeq
1$, are a typical realization of a homogeneous Gaussian random field,
Grishchuk and Zel'dovich showed~\cite{GZ} that they contribute to the
amplitude of the quadrupole anisotropy of the cosmic microwave
background (CMB) as $Q \simeq (L H_0)^{-2}$. By constraining $Q_{\rm
rms} \lsim 2\times10^{-5}$ from COBE~\cite{COBE}, one can see that the
size of the homogeneous patch should be at least $L_0 \gsim 500
H_0^{-1}$~\cite{KTF}. This requires $N_e \gsim 70$ or $|1-\Omega_0|
\lsim 10^{-4}$. Thus in standard inflation, an open universe with
$\Omega_0$ significantly less than one is incompatible with the large
scale homogeneity we observe in the CMB.

\section{MODELS OF OPEN INFLATION}

Open inflation solves the homogeneity problem by inflating the
universe in a false vacuum and then creating a bubble within which
our patch of the universe expanded to `almost' flatness, thanks to the
energy density of an inflaton field. According to this picture we live
inside a bubble that nucleated from de Sitter space by quantum
tunneling with an extremely small probability. This ensures two
things, first that there will be no collisions with other bubbles, at
least in our past light cone, and second that the nucleated bubble is
extremely spherically symmetric, defining a very homogeneous initial
hypersurface. The bubble walls then expand at the speed of light,
while the space-time within the bubble becomes that of an open
universe~\cite{CDL}, with infinite equal-time hypersurfaces.

The first models of open inflation~\cite{BGT,STY} considered a single
scalar field trapped in a metastable state that later tunneled to the
true vacuum with a non-zero energy density. The field then rolled down
a very flat potential, inflating the required amount of $e$-folds to
give an open universe. Note that in order to produce a universe with
say $\Omega_0 = 0.3$, we have to fine-tune the number of $e$-folds to
within one percent, which is a rather mild fine-tuning. These models,
though, had the unpleasant feature of strongly contrived potentials for
the inflaton field. In order to tunnel without producing too large
inhomogeneities on large scales, we need a large mass in the false
vacuum. One of the dangers of quantum tunneling with a small mass is
the existence of the Hawking-Moss instanton~\cite{HM}. In this case,
the field jumps to the top of the barrier between the two vacua and
very slowly `rolls down' the potential. Very large amplitude quantum
fluctuations are then produced that are not inflated away and would
unacceptably distort the observed anisotropy of the CMB. For that
reason alone it is assumed that the mass of the tunneling field should
be much larger than the rate of expansion at the false vacuum. On the
other hand, a very small mass of the inflaton field is required after
tunneling to give the observed amplitude of density perturbations in
the CMB. Linde and Mezhlumian~\cite{LM} suggested a simple way out by
including two fields, one with a large mass, responsible for
tunneling, and the other with a very small mass, responsible for
inflation inside the bubble. Except for the finite number of
$e$-folds, the second phase of open inflation is identical to standard
inflation. The scalar field ends inflation by oscillating at the
bottom of the potential and releasing its potential energy density
into radiation. The standard hot big bang cosmology follows
thereafter.

\section{DENSITY PERTURBATIONS}

If open inflation is to be a good model of the universe it should not
only solve the flatness and homogeneity problems but also account for
the temperature fluctuations we observe in the CMB. There are in
principle two sources of curvature perturbations in open inflation,
the quantum fluctuations of the inflaton field that are stretched to
cosmological scales by the expansion, and quantum fluctuations in
the bubble wall produced during the bubble nucleation. The former have
been thoroughly studied in recent papers~\cite{BGT,STY,LW}. Their main
result is the existence, in the spectrum of a scalar field with $m^2 <
2H^2$ in the false vacuum, of a single discrete supercurvature mode
that has to be taken into account together with the subcurvature
modes. In single field open inflation~\cite{BGT,STY}, we have seen
that such a small mass is incompatible with observations, but in
two-field models~\cite{LM} the inflaton field could indeed have a very
small mass in the false vacuum. This means that the universe could
actually be in a process of self-reproduction, and thus extremely
inhomogeneous~\cite{Book}. In that case, very large scale density
perturbations could affect the amplitude of the lower multipoles of
the temperature anisotropies, as discussed in the introduction. This
is the so-called Grishchuk-Zel'dovich effect~\cite{GZ}. We have
recently evaluated this effect in the open universe case~\cite{OGZ}
and found strong constraints on the amplitude of very long wavelength
perturbations contributing to the first CMB multipoles. These
constraints can be used~\cite{JGB} to bound the mass of the inflaton
field in the false vacuum.

But there are also potentially dangerous curvature perturbations
arising from quantum fluctuations of the bubble wall at the moment of
tunneling. They have been addressed in a very qualitative way by Linde
and Mezhlumian~\cite{LM}, and later studied in detail in
Ref.~\cite{JGB}. Most of the results of Refs.~\cite{BGT,STY,LM} were
done in the thin wall approximation, which is valid for most
potentials with a deep false vacuum minimum and a large potential
barrier between the two vacua. They also assume the tunneling occurs
from de Sitter to Minkowski space-time. However, the new ingredient in
open inflation is precisely the non-zero energy density of the true
vacuum which could still drive inflation to almost flatness. The
instanton action associated with the more general quantum tunneling
process from de Sitter to de Sitter was computed long ago by
Parke~\cite{Parke}. It is possible to calculate the tunneling action
of open inflation, and then compute the amplitude of curvature
perturbations from quantum fluctuations in the bubble wall, following
the covariant formalism of Ref.~\cite{GV}.

The origin of curvature perturbations in the bubble wall can be
understood as follows. Due to a large instanton action, the
probability of tunneling $\Gamma \sim \exp(-S_E)$ is extremely small.
The radius of curvature of the bubble corresponds to an extremum of
the instanton action, and the amplitude of quantum fluctuations in
this radius can be obtained from the second derivative of the
instanton action. These fluctuations can be understood as long
wavelength homogeneous perturbations in the curvature of the bubble
wall. In the open de Sitter coordinates, the bubble wall is a
time-like hypersurface at a fixed radial coordinate $\sigma$ which
asymptotically determines a space-like hypersurface at a fixed
comoving time $\eta$ inside the bubble. Thus perturbations in the
radius of the bubble propagate inside as perturbations in the time it
takes to end inflation~\cite{JGB}. This generates curvature
perturbations inside the bubble, very much like adiabatic density
perturbations from quantum fluctuations of the inflaton
field~\cite{GP}. One can show~\cite{JGB} that for most open inflation
models, due to the gravitational effects at bubble nucleation, the
amplitude of bubble wall fluctuations can be much smaller than those
of the inflaton field in the subsequent phase of inflation inside the
bubble.

The present models of open inflation seem to work quite well with very
reasonable parameters, at least as reasonable as those of standard
inflation. A different issue is whether these models will turn out to
be the correct description of the origin of our patch of the universe.
Fortunately, cosmology has become a science and within a few years we
will be able to tell, from the shape and amplitude of the spectrum of
density perturbations in the cosmic microwave background, whether our
patch of the universe is indeed open or flat~\cite{CMB}. In any case,
it is encouraging to see that the inflationary paradigm is able to
accommodate an open universe, even if we never have to make use of it.

\end{document}